\documentclass[12pt]{iopart}
\usepackage{graphicx,epsfig}
\usepackage{amssymb}
\usepackage{iopams}
\usepackage{float}

\begin{document}
\title[Primordial black hole formation]{Primordial blackhole formation: Exploring chaotic potential with a sharp step via the GLMS perspective}

\author{Rinsy Thomas$^{1,2}$\footnote{Corresponding author: Rinsy Thomas, rinsy@cmscollege.ac.in}, Jobil Thomas$^3$, Minu Joy$^4$}
\address{$^1$ School of Pure and Applied Physics, Mahatma Gandhi University, Kottayam,  686560, Kerala, India}
\address{$^2$ Department of Physics, CMS College, Kottayam, 686001, Kerala, India }
\address{$^3$ Department of Physics, Indian Institute of Technology Hyderabad, 502285, Telangana, India}
\address{$^4$ Department of Physics, Alphonsa College, Pala, 686574, Kerala, India}
\ead{rinsy@cmscollege.ac.in, jobilthomas1996@gmail.com, minujoy@alphonsacollege.in}

\begin{abstract}
A sharp step on a chaotic potential can enhance primordial curvature fluctuations on smaller scales to the $\mathcal{O}(10^{-2})$ to form primordial black holes (PBHs). The present study discusses an inflationary potential with a sharp step that results in the formation of PBHs in four distinct mass ranges. Also this inflationary model allows the separate consideration of observable parameters $n_s$ and $r$ on the CMB scale from the physics at small scales, where PBHs formation occur. In this work we computed the fractional abundance of PBHs ($f_{PBH}$) using the GLMS approximation of peak theory and also the Press-Schechter (PS) formalism. In the two typical mass windows, $10^{-13}M_\odot$ and $10^{-11}M_\odot$, $f_{PBH}$ calculated using the GLMS approximation is nearly equal to 1 and that calculated via PS is of $10^{-3}$. In the other two mass windows $1M_\odot$ and $6M_\odot$, $f_{PBH}$ obtained using GLMS approximation is 0.01 and 0.001 respectively, while $f_{PBH}$ calculated via PS formalism yields $10^{-5}$ and $10^{-6}$. The results obtained via GLMS approximation are found to be consistent with observational constraints. A comparative analysis of $f_{PBH}$ obtained using the GLMS perspective and the PS formalism is also included. 

\end{abstract}



\section{Introduction}
The detection of gravitational waves (GWs) \cite{1979ZhPmR..30..719S} from the merger of binary black holes marked a significant milestone in the field of astrophysics and marked the beginning of the era of multimessenger astronomy \cite{PhysRevLett.116.061102}. GWs are ripples in space-time that are produced by accelerating massive objects, such as binary black hole systems. These GWs \cite{einstein1918gravitational} were first predicted by Albert Einstein in his general theory of relativity in 1915 \cite{einstein1915general}. However, it took decades of technological advancements to detect these subtle distortions in space-time. The advent of gravitational wave (GW) detectors like the Laser-Interferometer GW Observatory (LIGO) and Virgo paved a new way to observe these events. The first detection of GWs occurred on September 14, 2015, by the LIGO observatories. This signal was generated by the merger of two binary black holes located around 1.3 billion light years away \cite{PhysRevLett.116.061102}. These GWs travel freely through the universe once generated and act as a powerful tool to explore the early universe.  Various potential sources for the origin of GWs are investigated, including reheating after inflation \cite{PhysRevD.56.653, PhysRevLett.120.031301, PhysRevD.97.023516}, phase transitions \cite{PhysRevD.47.4372, PhysRevD.49.2837}, topological defects \cite{PhysRevD.23.852, PhysRevD.31.3052}, etc. Moreover, GWs originating from various distinct sources are uncorrelated, leading to the generation of a stochastic GW background. Consequently, the analysis of GW signals observed by various pulsar timing array (PTA) experiments, such as NANOGrav \cite{Agazie_2023}, EPTA \cite{Antoniadis_2023}, InPTA, PPTA and CPTA \cite{Agazie_2024} in the nHz frequency range pointed towards alternative cosmological explanations like GWs generated by cosmic strings or PBHs \cite{BLANCOPILLADO2018392}. These signals, believed to have originated during the inflationary phase \cite{Star, PhysRevD.23.347, PhysRevD.32.1899} of the universe, require further clarification through future PTA observations \cite{vagnozzi2023inflationary}. Additionally, incorporating data from GW standard sirens into existing cosmic models enhances precision in determining the interaction strength between dark matter and dark energy \cite{Yang_2019}. 

In the context of inflationary cosmology, if the scalar perturbations are large enough on small scales, this can result in the production of abundant PBHs. This situation could arise if the inflationary perturbation spectrum showed a non-Gaussianity and substantial blue tilt \cite{Young_2022}, or alternatively if the inflaton field experienced a slower roll for a particular duration of time that was much shorter than the entire inflationary phase \cite{PhysRevD.40.1753, polarski1992spectra, Starobinsky:1992ts, PhysRevD.54.6040}. Due to Hawking radiation, PBHs with a mass smaller than approximately $5\times 10^{-19}M_\odot$ have already undergone evaporation. However, PBHs with a mass greater than this threshold can remain in stable existence to the present day \cite{Hawking1975}. The PBHs formed during the early epoch of our universe could have significant implications \cite{Zhao2022BayesianIF, Cang_2023}, as they might seed the formation of supermassive black holes in galactic nuclei and AGN's \cite{10.1093/mnras/206.4.801, PhysRevD.66.063505}, influence the ionization history of the universe \cite{Afshordi_2003, Belotsky_2015} and contribute to the overall density of dark matter \cite{CHAPLINE1975, Prime}. The abundances of PBHs denoted as $f_{PBH}$ is characterised by its fraction within the current dark matter content. When $f_{PBH}\approx0.1$, PBHs become a plausible candidate for dark matter \cite{PhysRevD.96.123523, PhysRevD.98.023536}. If $f_{PBH}<<10^{-3}$ their potential as a dark matter candidate within the specified mass range can be excluded.

Furthermore, recent studies have shown that, considering the quantum effects such as the memory burden effect, semiclassical evaporation constraints on PBHs are altered and this effect slows down the PBHs evaporation, thereby allowing those with masses below $10^9 g$ to survive until the present day and significantly contribute to dark matter \cite{alexandre2024new, thoss2024breakdown}. Thus the ultralight PBHs with masses below $10^9$g can have significant implications for dark matter content and GW phenomology. PBHs with masses $10^9$g evaporate before big-bang nucleosynthesis (BBN) and can temporarily dominate the energy density of universe, leading to significant small scale density fluctuations and can induce stochastic gravitational wave background \cite{Papanikolaou_2021, Domènech_2021}. Also the memory burden effect, where Hawking evaporation slows after a PBH has lost about half its mass, extends the PBH's lifetime and enhances the GW signal \cite{Papanikolaou_2022,balaji2024probing}.

Within the framework of single field inflationary models, the formation of PBHs is plausible when the potential exhibits characteristics like a nearby inflection point or a saddle-like region. This feature slows the motion of the inflaton field, resulting in an intensified peak within the perturbation spectrum \cite{PhysRevD.50.7173, PhysRevD.96.063503, PhysRevD.97.023501, Bhaumik_2020}. In the present work, we consider a small step-like feature in the base inflationary potential $V_b(\phi)$. This step effectively acts like a speed breaker by locally slowing the scalar field motion. Consequently, this leads to a sharp increase in the power spectrum at least to the $\mathcal{O}(10^{-2})$ and inducing stochastic GW background. Thus, the inherent localised nature of the speed breaker mechanism allows the generation of PBHs across a broad spectrum of masses spanning from the extremely light weight, $10^{-17}M_\odot$ to the immensely massive $10^2M_\odot$. Note also that this inflationary model with sharp step can produce ultralight PBHs which have significant contribution to the totality of dark matter and are associated with a rich GW phenomology. Remarkably, this inflationary mechanism allows for a wide range of PBH masses without significantly changing $n_s$ and $r$ on the cosmic microwave background (CMB) scales. We calculate the $f_{PBH}$ using an approximate method of peak theory (GLMS approximation) \cite{PhysRevD.70.041502} and Press-Scheter (PS) theory \cite{1974ApJ...187..425P} and then compare the results.

The structure of our article is as follows. In Section 2 we consider the chaotic inflationary model featuring a step in its potential, which has a brief period of ultra-slow roll inflation. The potential parameters are selected such that the feature enhances the curvature perturbations at small scale, which is essential for PBHs formation, with out affecting the key observational inflationary parameters like scalar spectral index $n_s$ and tensor to scalar ratio $r$ on CMB scales. The production of PBH and its mass are discussed in Section 3. Section 4 covers the fractional abundance of PBHs due to this chaotic inflationary model with sharp step, for four distinct mass windows of PBHs using the GLMS approximation of the peak theory formalism and also in the PS formalism. In Section 5, the summary and conclusions are discussed.

\section{Chaotic Inflationary model with step}
The exploration into the spectrum of adiabatic perturbations within the universe, particularly in scenarios where singularities are present in the inflation potential, holds notable importance in the realm of early universe cosmology \cite{Starobinsky:1992ts, PhysRevD.64.123514, PhysRevD.76.023503, JanHamann_2010}. In order to achieve relevant abundances of PBHs \cite{PhysRevD.96.063503, Germani2017OnPB}, from a localised perturbation in inflaton potential, we consider a chaotic inflationary potential with sharp step given by equation ($\ref{eq:3}$). The step acts as a speed-restraining element by slowing down the motion of the scalar field. The step in the inflationary potential is achieved by the hyperbolic tangent function which is distinguished by its characteristics including its height $c$, width $d$ and position $\phi_{step}$.

\begin{equation} \label{eq:3}
{
 V(\phi) = \frac{1}{2}m^2\phi^2\left[1+c \tanh\left(\frac{\phi-\phi_{step}}{d}\right)\right]
 } 
\end{equation}
 The occurrence of step in the chaotic inflationary potential \cite{PhysRevD.64.123514} can be explained based on various fundamental physical theories especially symmetry-breaking phase transitions, supergravity theories, effective field theory and string theory  in high-energy physics framework. Symmetry-breaking phase transitions \cite{PhysRevD.97.115025, GarciaBellido2002, Pallis2014} can induce features in the inflation potential due to localised changes in the potential energy landscape. In supergravity theories, the interactions between moduli fields and the inflaton naturally generate features in inflation potential \cite{Kallosh2010}. In general sharp step in the chaotic potential can produce different features in the primordial power spectrum depending on the potential parameters $c$, $\phi_{step}$ and $d$. The sharp step results in localised oscillations in the primordial power spectrum \cite{PhysRevD.64.123514,  Thomas2023101313} and these oscillations are typically more pronounced at larger scales if the step is encountered earlier during inflation when these modes are exiting the horizon. Also the step in the potential manifest themselves as distinct characteristics within the angular power spectrum of CMB \cite{Hazra_2010}. Several studies focused on a model-independent reconstruction of the primordial spectrum from the CMB observations have indicated the existence of specific features in the spectrum \cite{Chandra_2021}. The angular power spectrum of the CMB exhibits notable characteristics such as suppressions near the multipole moment $\ell = 2$, as well as a dip and a bump around $\ell \approx 2$ and $\ell \approx 40$, respectively \cite{PhysRevD.75.123502}. Also, distinct features are observed at $\ell \approx 300$, between $\ell \approx 750$ and $\ell \approx 850$, and in the range of $\ell \approx 1800$ to $\ell \approx 2000$ \cite{Hazra_2014}. If the sharp step is positioned such that the inflaton field encounters it later during inflation, the primordial power spectrum can exhibit a pronounced peak at smaller scales. However, PBH formation is influenced by small-scale features in the primordial power spectrum.
 
\begin{figure}[!ht]
  \centering
  \includegraphics[width=1\textwidth]{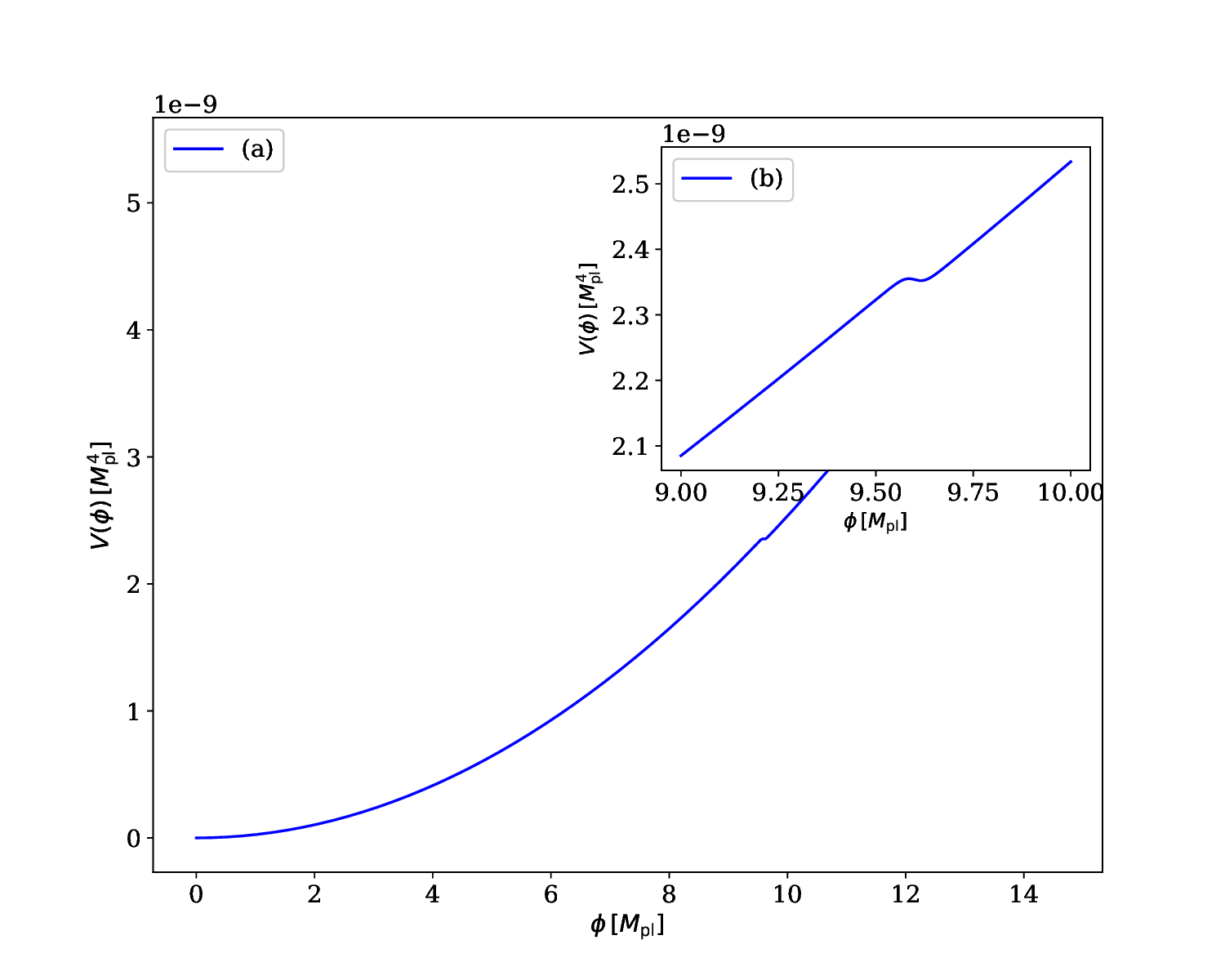}
  \caption{Plot of the chaotic inflationary potential with sharp step and enlarged inset showcasing the feature.}
  \label{Fig:3.1}
\end{figure}

The inflationary model in our work also has the advantage that it allows the separate consideration of observable parameters $n_s$ and $r$ on the CMB scale from the physics at small scales, where PBH formation occur. It is evident from equation (\ref{eq:3}) that potential $V(\phi)$ is characterised by four parameters $\{m, c, \phi_{step}, d\}$. However, since the inflaton mass, $m$ determines the overall CMB normalization, only the parameters $\{c, \phi_{step}, d\}$  are relevant when considering the PBH formation. In the present study we use the natural system of units where the reduced planck mass $M_{pl}=1$. The scalar field $\phi$, value of $\phi$ at which step occurs in the potential $\phi_{step}$ and the width parameter $d$ are expressed in units of $M_{pl}$ while the parameter $c$ is dimensionless. $V(\phi)$ has units of $M_{pl}^4$ maintaining dimensional consistency within the natural units framework. Figure \ref{Fig:3.1} showcases the potential V($\phi$), where the main plot (a) effectively presents the potential across the entire range of $\phi$, capturing the overall behaviour of the function. However, due to the presence of a minor but significant feature (step) in the potential, we employ an inset (b) to provide a magnified view of the region where the feature is present.

In the context of single-field inflation, the driving force behind inflation is a minimally coupled canonical scalar field with an appropriate potential V($\phi$) and the corresponding action is,
\begin{equation}
    \mathcal{S} = \int d^4x \sqrt{-g}\left[\frac{M_{pl}^2}{2}{R}-\frac{1}{2} \partial_{\mu}\phi \partial^{\mu}\phi-V(\phi)\right]
\end{equation}
where ${R}$ is the Ricci scalar and $M_{pl}$ is the reduced Planck mass.\\
For a spatially flat universe, the equation of motion of the scalar field and the scale factor is governed by,
\begin{equation}
    H^2 = \frac{1}{3M_{pl}^2}\left[V(\phi)+\frac{1}{2}\dot \phi^2\right]
\end{equation}
where Hubble parameter, $H=\frac{\dot a}{a}$.
The extend of inflation is given by the amount of e folds during inflation,
\begin{equation}
    N_e=\ln{\frac{a(t_{end})}{a(t_0)}}
\end{equation}
where $t_{0}$ and $t_{end}$ corresponds to the time at the beginning and the end of inflation. $N_e$ decreases to zero at the end of inflation.

The standard technique for analysing inflation is the slow-roll approximation. In the standard slow-roll inflation, values of the slowroll parameters $\epsilon=\frac{1}{2}M_{pl}^2\left(\frac{V^\prime}{V}\right)^2$ and $\eta=M_{pl}^2\left(\frac{V^{\prime\prime}}{V}\right)$ are typically significantly smaller than 1. However, during the ultra-slow roll stage, there is significant deviation from the standard slow roll conditions. In order to analyse this, we study the Hubble flow parameters $\epsilon_1\approx\epsilon$ and $\epsilon_2\equiv\frac{\dot{\epsilon_1}}{\epsilon_1 H}=-2\eta+4\epsilon_1$. The step feature in the potential locally modifies the inflaton field dynamics, causing temporary deviations from slow-roll conditions.

In the slow roll scalar field, an important equation governing the scalar curvature perturbation ($\mathcal{R}$) is the Mukhanov-Sasaki equation \cite{10.1143/PTP.76.1036, Mukhanov:1988jd}.
The equation that governs the motion of Fourier components, denoted as $u_k$ is as follows,
\begin{equation}\label{eq:Pk}
    u_k^{\prime\prime} + \left(k^2-\frac{z^{\prime\prime}}{z}\right)u_k=0\;.
\end{equation}
Here the prime symbol indicates the differentiation with respect to conformal time, $k$ is the modulus of the wave number.\\

The scalar power spectrum $\mathcal{P}(k)$ is usually defined as
\begin{equation}\label{eq:P}
    \mathcal{P}(k)=\sqrt{\frac{k^3}{2\pi^2}}\left|{\frac{u_k}{z}}\right|^2_{k\ll aH}\; .
\end{equation}
 Similarly, the equation that governs the generation of gravitational wave modes by tensor perturbations$(\psi)$ \cite{1979ZhPmR..30..719S} during inflation is,
 \begin{equation}
    v_k^{\prime\prime} + \left(k^2-\frac{a^{\prime\prime}}{a}\right)v_k=0
 \end{equation}
 where $v_k=a\psi_k$.
 The power spectrum $\mathcal{P_T}(k)$ for gravitational waves follows an analogous form to equation (\ref{eq:P}) that is,
 \begin{equation}
     \mathcal{P_T}(k)=\frac{k^3}{2\pi^2}\left|\frac{v_k}{a}\right|^2 \; .
 \end{equation}
 However, the potential in the present work exhibits a distinct step-like feature and as a result 
 $V^{\prime}(\phi)$ and $\dot{\phi}$ need not be small. As a consequence of these unique characteristics, we opt to numerically solve the complete mode equation by the Runge-Kutta method without resorting to any approximations beyond those inherent in the framework of perturbation theory. In the context of our analysis it is essential to consider the observable parameters, scalar spectral index $n_s$ and tensor to scalar ratio $r$. The step like feature in the present inflationary model is positioned in such a way that it leads to a brief period of ultra-slow roll inflation, enhancing the power spectrum of curvature perturbation at small scales, which is important for PBH formation without affecting $n_s$ and $r$ on CMB scale.
 
 The presence of a step-like feature in the potential is required on a smaller scale $k\gg k_*$, that results in an amplification of perturbations leading to large enhancement in the scalar power spectrum while the enhancements are suppressed in the tensor power spectrum \cite{PhysRevD.95.083519}. The enhancement in the scalar power spectrum leads to the generation of PBHs. 
 Our focus in this section is exclusively on the CMB scales $10^{-4}Mpc^{-1}$ to $1Mpc^{-1}$, with the aim of assessing how well the observable parameters are consistent with the Planck data. In Table \ref{tab:my_label} the four sets of parameter values $\{c, d, \phi_{step}\}$ have been chosen to ensure that the values $n_s$ and $r$ on the pivot scale of the CMB $(k_*=0.05Mpc^{-1})$ are consistent with the Planck data, while also facilitating PBH production.

\begin{table}[h!]
\centering
\renewcommand{\arraystretch}{1.5}
\begin{tabular}{@{}ccccccc}
\hline
\noalign{\vskip\doublerulesep}
\hline
Set & $c$ & $d M_{pl}$ & $\phi_{step}M_{pl}$ & $n_s$ & $r$\cr
\hline
 1& -7.91501$\times10^{-3}$ & 0.029 & 9.6 &  &  \cr
2& -6.97333$\times10^{-3}$ & 0.027 & 10.2 & 0.96 & 0.02 \cr
 3& -2.70720$\times10^{-3}$ & 0.003 & 12.2 &  &  \cr
 4& -2.52303$\times10^{-3}$ & 0.002 & 12.32 &  &  \cr 
\hline
\noalign{\vskip\doublerulesep}
\hline
\end{tabular}
\caption{Typical parameter values for the potential enabling PBH production, consistent with CMB data}
\label{tab:my_label}
\end{table}

\begin{figure}
    \centering
    \includegraphics[width=\textwidth]{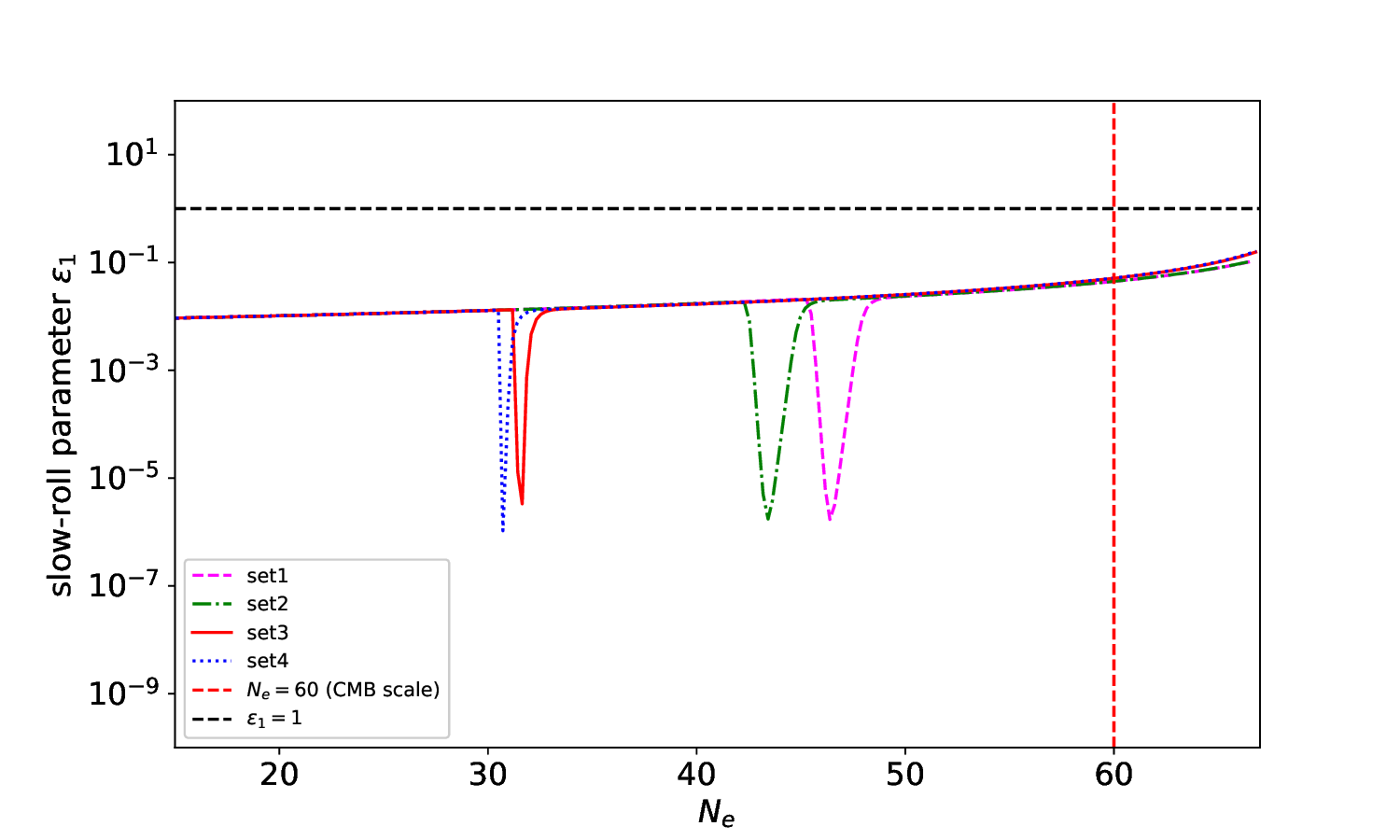}
    \caption{Slow roll parameter $\epsilon_1$ plotted as a function of number of e-folds for the chaotic inflationary potential with sharp step}
    \label{e1}
\end{figure}

\begin{figure}
    \centering
    \includegraphics[width=\textwidth]{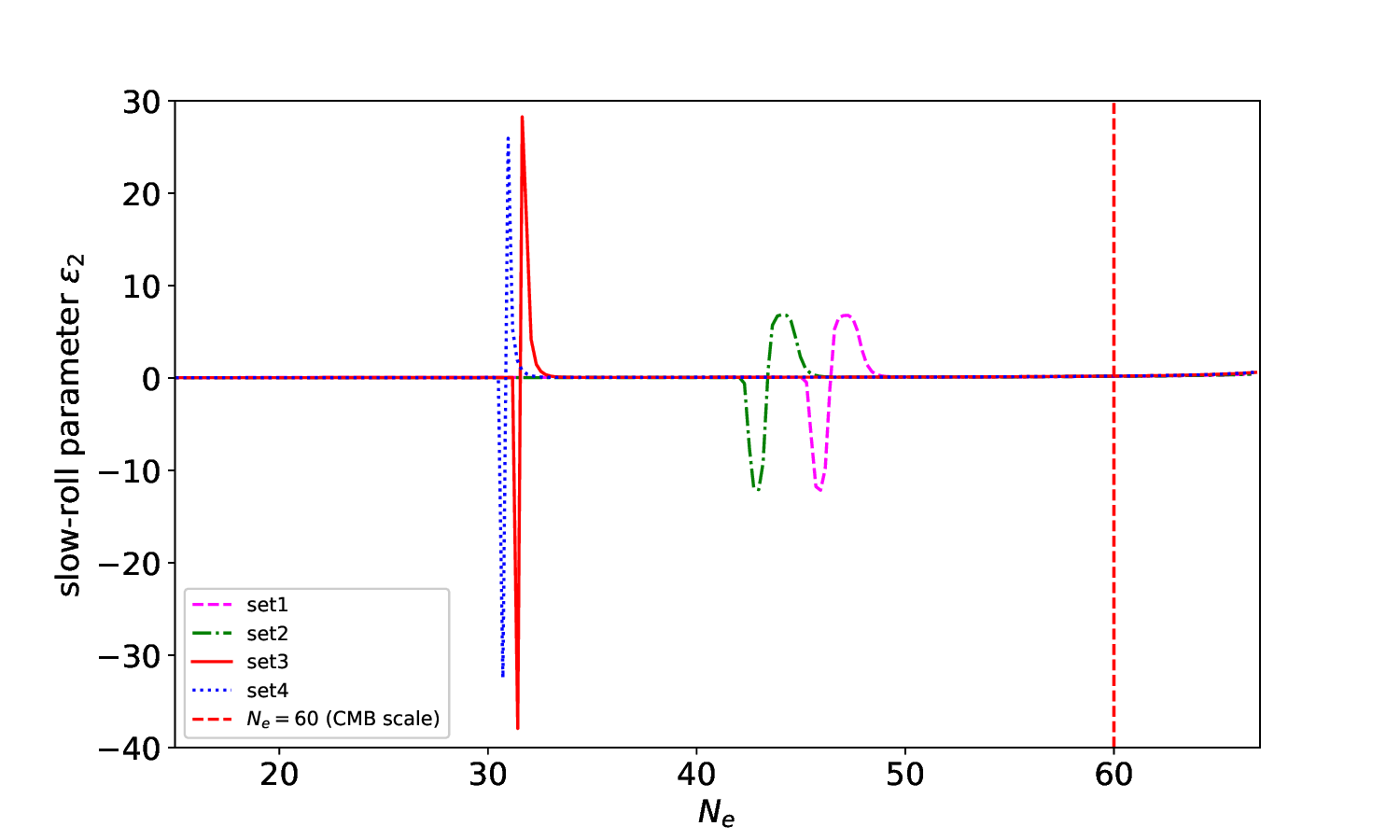}
    \caption{Slow roll parameter $\epsilon_2$ plotted as a function of number of e-folds for the chaotic inflationary potential with sharp step}
    \label{e2}
\end{figure}

The step feature modifies the potential locally and then the  inflaton field dynamics quickly settle back into a stable slow-roll trajectory. This behavior confirms the robustness of the attractor solution in our model. Figures \ref{e1} and \ref{e2} shows the local violation of slowroll conditions due to the presence of a step feature in the chaotic inflationary potential. The parameter $\epsilon_1$ remains small while $\epsilon_2$ can approach values around $\mathcal{O}(1)$ or higher \cite{dimopoulos2017ultra}, leading to significant impacts on the abundances of PBHs.


\begin{figure}[h!]
\centering
\includegraphics[width=0.9\textwidth]{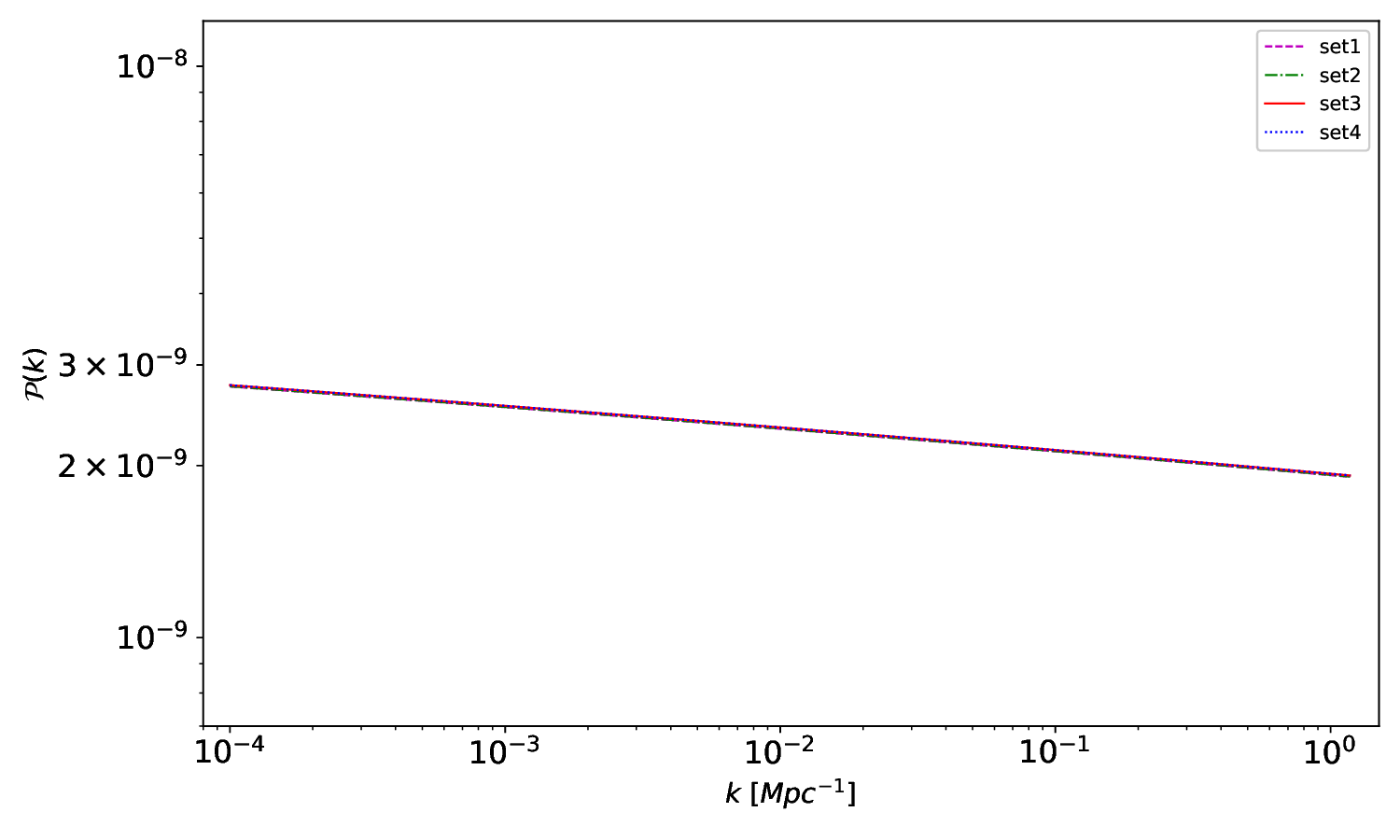}
\caption{Scalar power spectrum $\mathcal{P}(k)$ on large scale for the chaotic inflationary model with step, for four distinct parameter sets}
\label{Fig:3.2}
\end{figure}

\begin{figure}[h!]
\centering
\includegraphics[width=0.9\textwidth]{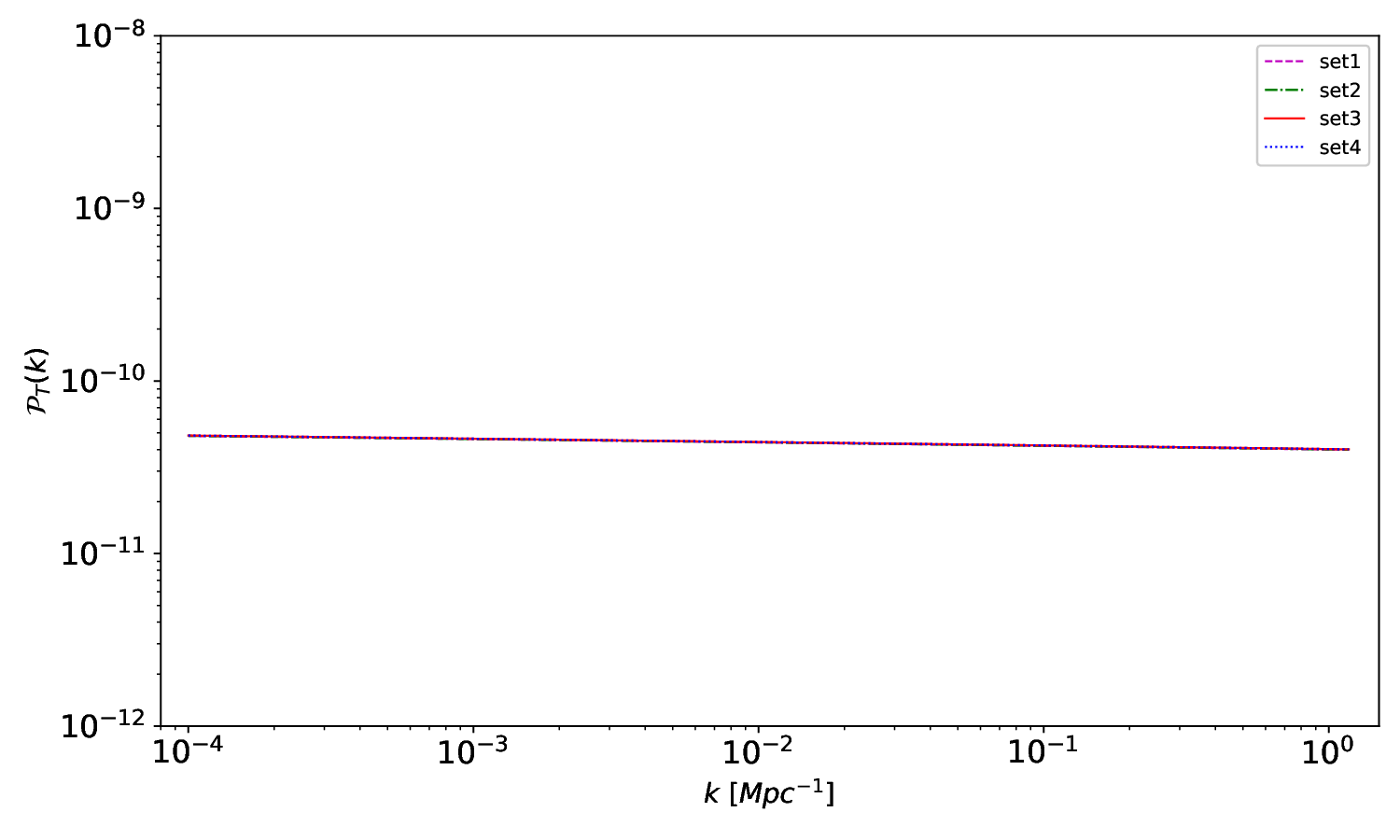}
\caption{Tensor power spectrum $\mathcal{P}_T(k)$ for the chaotic inflationary model with step, for four distinct parameter sets}
\label{Fig:3.2.b}
\end{figure}

Our analysis reveals remarkable uniformity in the four sets of parameter values. For all parameter sets, consistently same values are obtained for the spectral index $n_s$ and also for the tensor to scalar ratio $r$ at the CMB pivot scale. This highlights strong and stable characteristics in the predictions by the model. The scalar power spectrum $\mathcal{P}(k)$ and the tensor power spectrum $\mathcal{P}_T(k)$ on the CMB scale are shown in Figures \ref{Fig:3.2} and \ref{Fig:3.2.b} respectively. It is obvious that the scalar and tensor power spectra are nearly scale invariant on CMB scales for all the four sets.
 
\begin{figure}
     \centering
     \includegraphics[width=\textwidth]{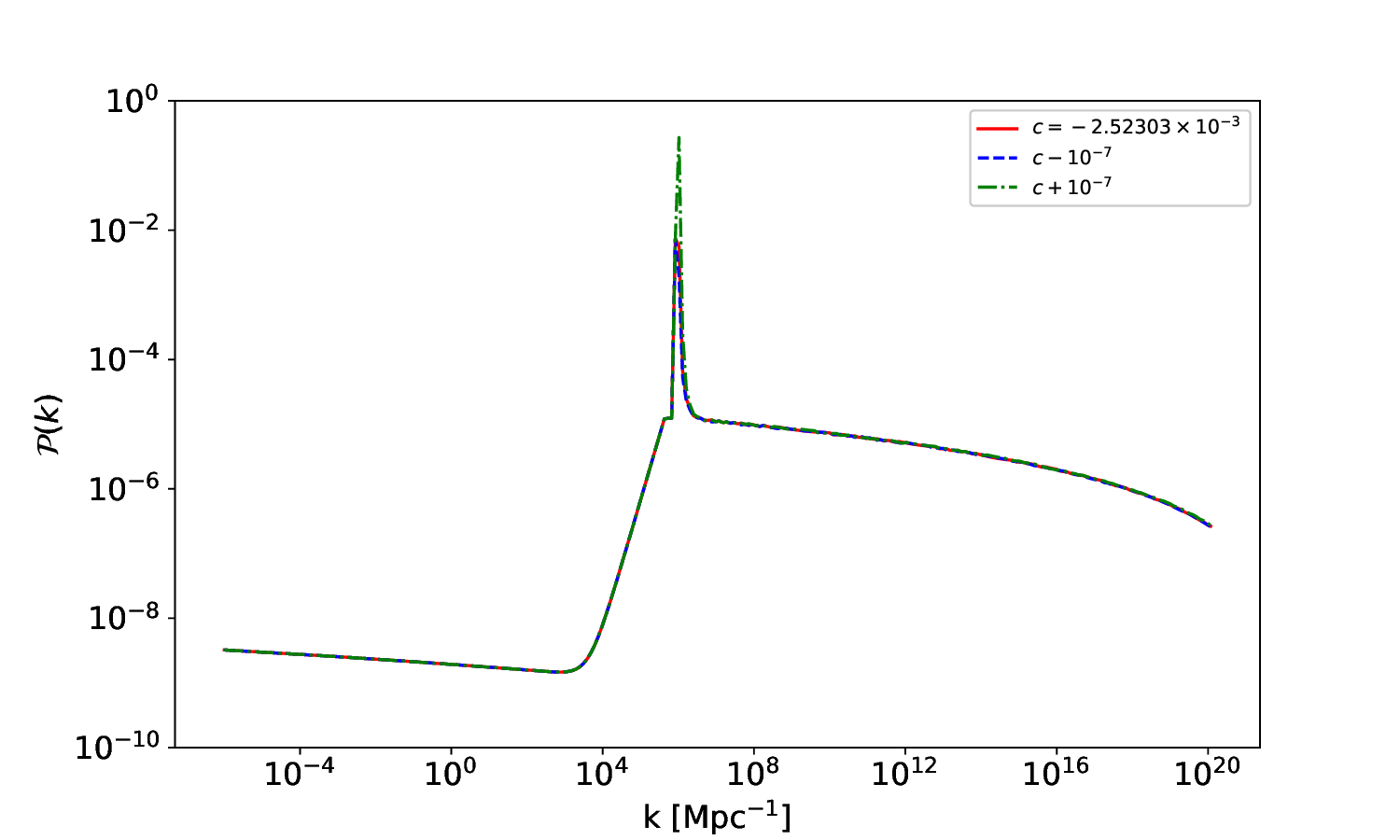}
     \caption{Sensitivity of scalar power spectrum to the potential parameter $c$ with $\phi_{step}=12.32$ and $d=0.002$ }
     \label{fig-sensitivity}
 \end{figure}

\begin{figure}
     \centering
     \includegraphics[width=\textwidth]{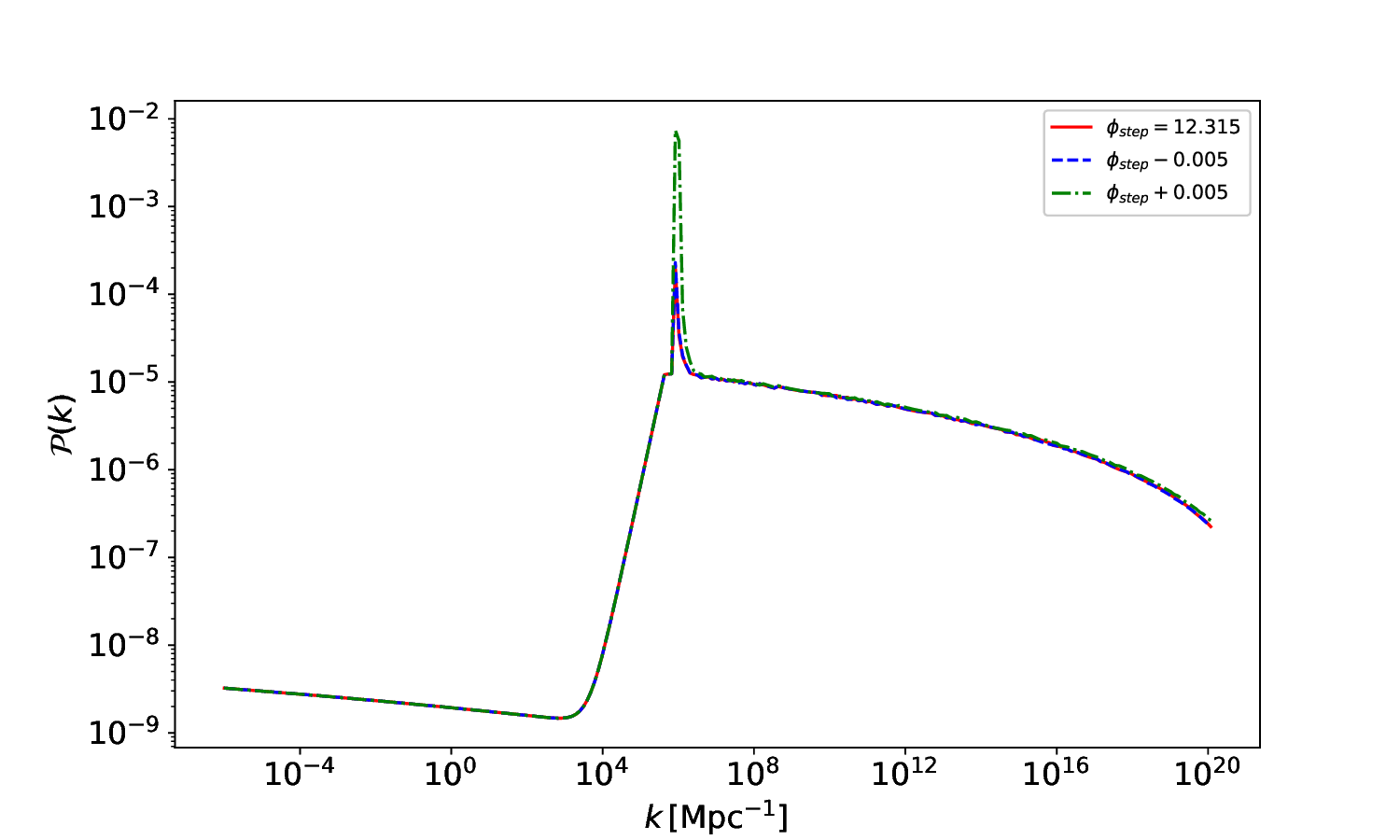}
     \caption{Sensitivity of scalar power spectrum to the potential parameter $\phi_{step}$ with $c=-2.52302\times10^{-3}$ and $d=0.002$. }
     \label{fig-sensitivity-phistep}
 \end{figure}

\begin{figure}
     \centering
     \includegraphics[width=0.9\textwidth]{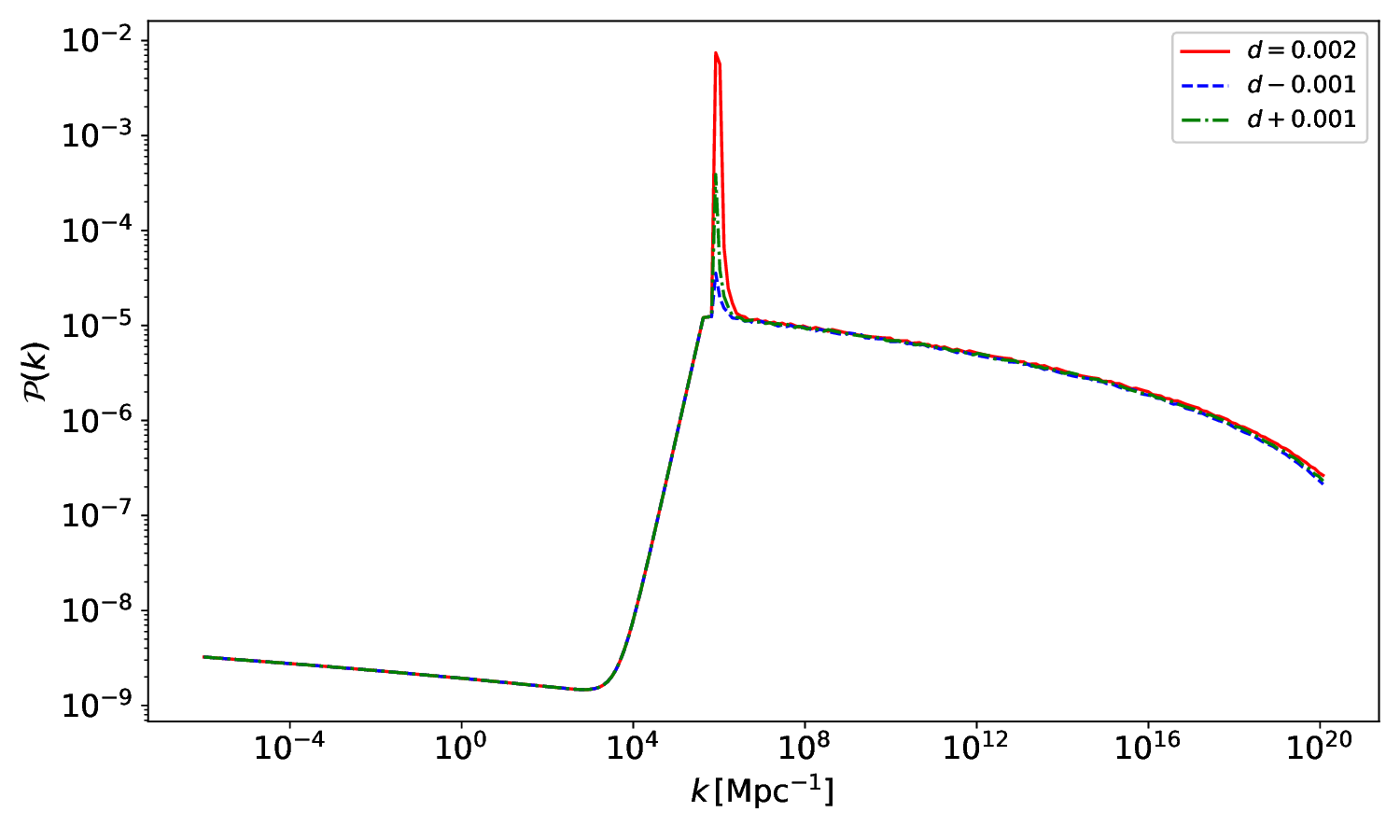}
     \caption{Sensitivity of scalar power spectrum to the potential parameter $d$ with $c=-2.52302\times10^{-3}$ and $\phi_{step}=12.32$. }
     \label{fig-sensitivity-d}
 \end{figure}

In this chaotic inflationary model with sharp step, fine tuning of the potential parameters is required, for the observables $n_s$ and $r$ to be consistent with CMB data and also to enhance primordial scalar power spectrum  at small scales at least to the $\mathcal{O}(10^{-2})$ for the formation of PBHs \cite{Cole_2023}. Hence a fine tuning of the parameter $c$ of the inflationary potential  by even a small fraction (one part in $10^7$) can significantly change the peak amplitude of the power spectrum. As shown in Figure \ref{fig-sensitivity}, for $c+10^{-7}$ just above $c=-2.52303\times10^{-3}$, peak value increases by a factor of $10^2$. However, a decrease in $c$ by the same amount $10^{-7}$ results in a minimal change in the peak amplitude, suggesting that $c$ is at or very near a threshold where a small increase results in a significant increase on the peak amplitude. Also, as it is clear from Figures \ref{fig-sensitivity-phistep} and \ref{fig-sensitivity-d}, fine tuning of $\phi_{step}$ up to two decimal places and $d$ up to three decimal places are enough to maintain the consistency of observables at CMB scale and to maintain the peak of scalar power spectrum at small scale where the PBHs formation occur.

\section{Primordial Blackhole production}
\begin{figure}[h!]
  \centering
  \includegraphics[width=0.9\textwidth]{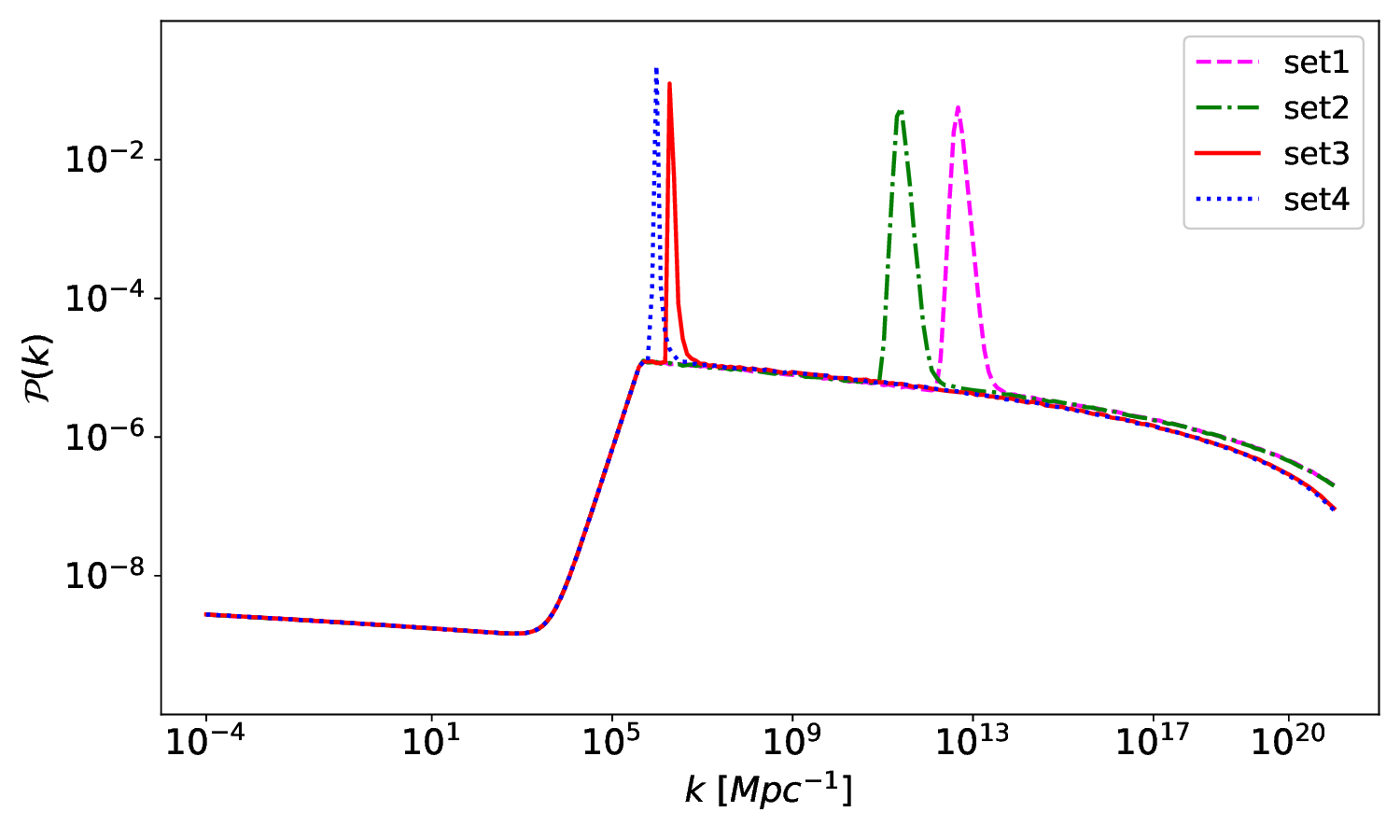}
  \caption{Scalar power spectrum $\mathcal{P}(k)$ on broader scales for four distinct parameter sets}
  \label{Fig:3.3}
\end{figure}

The presence of a distinctive feature in the inflationary potential on smaller scales leads to the formation of PBHs and by fine tuning the antisymmetric perturbation feature within the inflationary potential, it is possible to generate PBHs \cite{PhysRevD.96.063503, Bhaumik_2020, Germani2017OnPB, Gangopadhyay_2022, garcia-bellido2017primordial, Mishra_2020}
in specific mass ranges that could constitute all dark matter \cite{doi:10.1146/annurev-nucl-050520-125911}. As depicted in Figure \ref{Fig:3.1}, the distinctive step-like feature in the potential has the capacity to significantly slow the inflaton field, which is inherently undergoing a slow rolling motion. A significant decrease in $\dot{\phi}$ (with minimal change in the value of $H$) during the inflationary period results in $\epsilon$ decreasing appreciably from its pivot scale. This leads to a substantial enhancement in the amplitude of the scalar power spectrum. The observational parameters of the inflationary potential being consistent with large scale CMB observations, it is essential that $\mathcal{P}(k)$ undergoes a significant enhancement at least by a factor of $10^7$ within 40 e-folds of expansion. The generation of PBHs with higher masses necessitates the step to be smaller in height ($c$), and sharper in width ($d$). The $\mathcal{P}(k)$ calculated using the Mukhanov-Sasaki equation is plotted with respect to the wavenumber $k$ prior to the end of the inflationary period for four distinct sets of potential parameters $\{c,d,\phi_{step}\}$. The analysis reveals notable significance in the behavior of the scalar power spectrum on different scales, and it is shown in Figure \ref{Fig:3.3}.

We studied the tree-level calculations of the power spectrum for the formation of primordial black holes (PBHs). Recent studies \cite{Inomata_2023, PhysRevLett.132.221003, Choudhury_2023_QL, Choudhury_2023, choudhury2023nogo, ballesteros2024oneloop, franciolini2023loop} have examined the impact of small-scale one-loop corrections on the large-scale curvature power spectrum. However, a detailed analysis of one-loop corrections and the application of EFT \cite{Choudhury_2023_QL, Choudhury_2023} will be pursued in future work to significantly address the backreaction issue.

In the Carr–Hawking collapse model \cite{10.1093/mnras/168.2.399} the mass of a PBH formed during a specific epoch in the radiation dominated era, originating from the Hubble reentry of a significant fluctuation mode $k_{PBH}$ is related to the Hubble mass during its formation \cite{PhysRevD.96.063503, Sasaki_2018} and is expressed as follows,
\begin{equation}
    M_{PBH}=\gamma M_H = \gamma \frac{4\pi M_{pl}^2}{H} \; .
\end{equation}
Here $\gamma$ denoting the efficiency of collapse is taken to have a value of 0.2 for the formation of PBHs during the radiative epoch \cite{Sasaki_2018, 1975ApJ...201....1C, PhysRevD.96.043504}. In the radiative epoch \cite{PhysRevD.96.063503},
\begin{equation}
    H^2=\Omega_{0r}H_0^2\left(1+z\right)^4\left(\frac{g_*}{g_{0*}}\right)^{-\frac{1}{3}}\left(\frac{g^s_{0*}}{g_{0*}}\right)^{\frac{4}{3}} \; .
\end{equation}
Here $g_{0*}$ and $g^s_{0*}$ represent the effective degrees of freedom for the energy and entropy, respectively, at the current epoch. Meanwhile, the relativistic degrees of freedom for the energy density during the radiation-dominated epoch, when PBHs are formed, are estimated to be approximately $g_*\approx 106.75$ \cite{1975ApJ...201....1C}, and the present-day radiation density parameter is characterised by $\Omega_{0r}h^2= 4.18\times 10^{-5}$. By employing the principle of conservation of entropy during the adiabatic expansion of universe \cite{PhysRevD.96.063503, PhysRevD.97.023501, Sasaki_2018, PhysRevD.96.043504,  Carr_2021},
\begin{equation}\label{eq:M}
    \frac{M}{M_\odot}=1.13\times 10^{15}\left(\frac{\gamma}{0.2}\right)\left(\frac{g_*}{106.75}\right)^{-\frac{1}{6}}\left(\frac{k_{*}}{k_{PBH}}\right)^{2}
\end{equation}
where the solar mass $M_{\odot}=1.99\times10^{30}$ kg and the CMB pivot scale $k_*=0.05Mpc^{-1}$ \cite{planck2018x}.
The equation ($\ref{eq:M}$) suggests that solarmass PBHs are formed when scalar modes with large fluctuation characterised by a comoving wavenumber $k_{PBH}\approx 10^7k_*$ entering Hubble radius. The mass of PBHs obtained for four distinct potential parameter sets are given in Table $\ref{tab:PBH_mass}$.

 \begin{table}[hbt!]
    \centering
    \renewcommand{\arraystretch}{1.5}
    \begin{tabular}{@{}ccccccc}
    \hline   
    \noalign{\vskip\doublerulesep}
    \hline
   Set& $c$ & $d M_{pl}$& $\phi_{step} M_{pl}$ & $\frac{M_{PBH}}{M_\odot}$  \\  
   \hline
   1& -7.91501$\times10^{-3}$ & 0.029 & 9.6 & $10^{-13}$  \\              
   2& -6.97333$\times10^{-3}$ & 0.027 & 10.2 & $10^{-11}$\\               
   3& -2.70720$\times10^{-3}$ & 0.003 & 12.2 & 1 \\                   
   4& -2.52303$\times10^{-3}$ & 0.002 & 12.32 & 6 \\  
    \hline
    \noalign{\vskip\doublerulesep}
    \hline
\end{tabular}
    \caption{Potential parameter values enabling PBH production and the PBH mass}
    \label{tab:PBH_mass}
\end{table}
Thus we have explored a broader parameter space, which shows how different sets of parameters leads to PBH formation across different mass ranges. Specifically, this model leads to the formation of PBHs of $10^{-13}M_\odot$, $10^{-11}M_\odot$, $1M_\odot$, $6M_\odot$ without affecting $n_s$ and $r$ values on CMB scale. In a recent work on the PBH production from single field inflation, PBH mass range was obtained to span from $10^{-18}M_\odot$ to $10^{-6}M_\odot$ \cite{Gangopadhyay_2022}. In another study \cite{PhysRevD.104.083546} of PBHs, the fractional abundance of  PBHs mass is determined using the GLMS approximation, focusing on mass windows such as \(10^{-17} M_\odot\), \(10^{-13} M_\odot\) and \(30 M_\odot\) and achieved a PBH abundance of \(f_{PBH} \sim 0.1\).
\section{Primordial blackhole abundance}
The formation of PBHs from large primordial overdensities is a topic of significant interest, especially given its implications for the nature of dark matter and the potential observational signatures that PBH could produce \cite{doi:10.1146/annurev-nucl-050520-125911}. The importance of fractional abundance $f_{PBH}$ lies in its capacity to quantify the significance of the contribution of PBHs to the overall dark matter density, a fundamental aspect in understanding the composition and evolution of the cosmos. Observational constraints on $f_{PBH}$ as a candidate for dark matter have been derived through various methods. These constraints arise from considerations related to Hawking radiation and PBH evaporation, the observation of black hole mergers by the LIGO and Virgo collaborations \cite{PhysRevD.96.123523, PhysRevD.98.023536}, impacts on CMB, the $Lyman-\alpha$ forest \cite{PhysRevLett.123.071102} and 21 cm cosmology \cite{PhysRevD.100.043540}. These diverse constraints provide upper limits on the fraction of PBHs relative to total dark matter, helping to refine our understanding of the PBH population. The fractional abundance of PBHs at the current epoch is defined as
\begin{equation}
    f_{PBH}^{tot}=\int\frac{d M_{PBH}}{M_{PBH}} f_{PBH}(M_{PBH})\; .
\end{equation}
The fractional abundance of PBHs of a certain mass $M_{PBH}$ at the present epoch is typically defined by,
\begin{equation}
  f_{PBH}(M_{PBH})  = \frac{\Omega_{0PBH}(M_{PBH})}{\Omega_{0DM}}
\end{equation}
where $\Omega_{0PBH}(M_{PBH})$ is the density parameter of PBHs of mass $M_{PBH}$ and $\Omega_{0DM}$ is the total density parameter of dark matter. Thus, $f_{PBH}(M_{PBH})$ essentially gives the fraction of total dark matter that is made up of PBHs of a specific mass $M_{PBH}$ \cite{Carr_2021}.
\begin{equation}
    f_{PBH}(M_{PBH})=1.68\times10^8\left(\frac{\gamma}{0.2}\right)^\frac{1}{2}\left(\frac{g_*}{106.75}\right)^{-\frac{1}{4}}\left(\frac{M_{PBH}}{M_\odot}\right)^{-\frac{1}{2}}\beta(M_{PBH})
\end{equation}
where $\beta(M_{PBH})$ is the mass fraction of PBHs at the time of its formation.
\begin{equation}
   \beta({M_{PBH})} =\frac{\rho_{PBH}}{\rho_{tot}}\bigg|_{formation} .
\end{equation}
Analysing the formation and abundances of PBHs in the early universe offers a unique window into the primordial density fluctuations.  During the radiation dominated era, on comoving scale, density contrast $\delta$ can be linearly related to $\mathcal{R}$ by the following expression \cite{PhysRevD.70.041502},
\begin{equation}
    \delta=\frac{4}{9}\left(\frac{k}{aH}\right)^2\mathcal{R}\; .
\end{equation}
The dimensionless power spectrum of the density contrast $\mathcal{P}_{\delta}$ is related to the primordial comoving curvature power spectrum $\mathcal{P}(k)$ as
\begin{equation}\label{P_delta}
    \mathcal{P}_{\delta}(k)=\frac{16}{81}\left(\frac{k}{aH}\right)^4\mathcal{P}(k)\;.
\end{equation}
The abundances of PBHs can be calculated through $\mathcal{P}_\delta(k)$. For this, it is necessary to initially smooth out the perturbation across a specific scale $\mathfrak{R}=\frac{1}{k_{PBH}}$. This is essential to circumvent the issues related to nondifferentiability and divergence at high values of $k$ within the radiation field. This smoothing procedure is achieved by incorporating a window function $W(k,\mathfrak{R})$ \cite{young2019primordial, Tokeshi_2020, PhysRevD.97.103528} within the Fourier space. The variance of the density contrast that is coarse-grained on a specific scale, $\mathfrak{R}=\frac{1}{k_{PBH}}=\frac{1}{(aH)_{PBH}}$ is expressed as,
\begin{equation}\label{sigma1}
    \sigma^2_\delta=\int\frac{dk}{k}\mathcal{P}_{\delta}(k)W^2(k,\mathfrak{R})\; .
\end{equation}
The choice of window function can significantly influence the statistical properties of the smoothed density field, thereby influencing the estimated PBHs abundances. Gaussian window function $W(k,\mathfrak{R})$ is used to smooth or coarse grain the original density contrast field $\delta$ on a certain comoving Hubble scale $\mathfrak{R}$. Here the Gaussian window function $W(k,\mathfrak{R})$ was chosen since it provides a smoother and more continuous suppression of contributions from large scales, avoiding the oscillations in the power spectrum that can arise from the sharp cutoff characteristics of other window functions such as top-hat function \cite{PhysRevD.70.041502, Young_2014}.
The expression for the Fourier transform of Gaussian window function is given by
\begin{equation}\label{W}
W(k,\mathfrak{R})=exp\left(-\frac{1}{2}k^2\mathfrak{R}^2\right)\; .
\end{equation}
$W(k,\mathfrak{R})$ play a pivotal role in this context acting as a filter that emphasizes fluctuations on scales close to $\mathfrak{R}$ and suppresses fluctuations on smaller scale. 
Using equations (\ref{P_delta}), (\ref{sigma1}), (\ref{W}) and the fact $\mathfrak{R}=\frac{1}{k_{PBH}}=\frac{1}{(aH)_{PBH}}$, we can obtain the expression for the variance of the density contrast $\sigma_{\delta}$ as,
\begin{equation}\label{sigma2}
    \sigma^2_{\delta}=\frac{16}{81}\int\frac{dk}{k}\left(\frac{k}{k_{PBH}}\right)^4\exp{\left(-\frac{k^2}{k^2_{PBH}}\right)}\mathcal{P}(k) \;.
\end{equation}
Additionally, the spectral moment corresponding to the $i^{th}$ order of smoothed density contrast is defined as
\begin{equation}
    \sigma_i^2=\int_0^\infty\frac{dk}{k}k^{2i}W^2(k,\mathfrak{R})\mathcal{P}_\delta(k)=\frac{16}{81}\int_0^\infty\frac{dk}{k}k^{2i}W^2(k,\mathfrak{R})\left(k\mathfrak{R}\right)^4\mathcal{P}(k)
\end{equation}
where $i=0, 1, 2, ...$ and $\sigma_0=\sigma_\delta$.

Thus it is obvious that the $\sigma_i$ depends upon the primordial scalar power spectrum $\mathcal{P}(k)$ and hence $f_{PBH}$ is very sensitive to the peak value of $\mathcal{P}(k)$.

One can calculate $\beta(M_{PBH})$ by employing various methodologies within the theoretical framework, including the GLMS approximation \cite{PhysRevD.70.041502} of peak theory, Press-Schechter (PS) formalism \cite{1974ApJ...187..425P}, peak theory \cite{1986ApJ...304...15B, PhysRevD.104.083546} and other approximations inherent to peak theory \cite{Young_2014, Gow_2021, 10.1093/ptep/pty120}. In this work, our focus is on the determination of $\beta(M_{PBH})$ using the GLMS approximation and the PS formalism.

\subsection{GLMS Approximation}
Green, Liddle, Malik, and Sasaki (GLMS) introduced a convenient approximation of peak theory \cite{PhysRevD.70.041502}. GLMS approximation is an approach that involves the computation of $\beta(M_{PBH})$ using the peak theory formalism. This framework introduces the primordial over-density condition in relation to the peak amplitude of a fluctuation mode as opposed to the average value employed in the PS theory. In the peak theory formalism, the peak amplitude of $\beta(M_{PBH})$ (and consequently of $f_{PBH}(M_{PBH})$) generally exhibits higher values. Within the framework of peak theory, the peak value denoted as the relative density contrast $\nu$ plays a pivotal role. This parameter is defined as the ratio of density fluctuation $\delta$ to its standard deviation $\sigma_{\delta}$; $\nu=\frac{\delta}{\sigma_{\delta}}$ and its threshold $\nu_{th}=\frac{\delta_{th}}{\sigma_{\delta}}$. It is important to note that $\nu_{th}$ is not a constant value, as it varies due to the dependence of $\sigma_{\delta}$ on the chosen smoothing scale $\mathfrak{R}$. The PBH formation threshold $\delta_{th}$ depends on the shape of the collapsing overdensity region . Thus it is obvious that $\delta_{th}$ depends not only on the amplitude of scalar power spectrum but also on the shape of these power spectrum \cite{PhysRevD.100.123524, PhysRevLett.122.141302}. Also the PBH formation in the early universe is influenced by the equation of state (EOS) at the time of formation and $\delta_{th}$ have a strong dependence on EOS \cite{PhysRevD.88.084051, Escrivà_2021, PhysRevD.105.124055}. $\delta_{th}$ depends on the primordial non-Gaussianities in the initial perturbation field which inturn affects the PBHs abundances \cite{Kehagias_2019}. In addition the critical threshold for PBH formation is not largely influenced by asphericities in the initial perturbation and it indicates that the PBH formation threshold remains largely unchanged \cite{PhysRevD.102.043526} but it is significantly influenced by anisotropies in the initial density perturbation \cite{PhysRevD.106.083017}. However, several numerical and analytical investigations have suggested the permissible range for $\delta_{th}$. In particular for the formation of PBH during the radiation dominated epoch, these studies have indicated that $\delta_{th}$ could range from 0.33 to as high as 0.66. $\delta_{th}$ is related to the equation of state of the background $w$ \cite{PhysRevD.88.084051, Niemeyer_1999, Musco_2005} and we studied nearly monochromatic PBH mass functions and adopted a threshold value of $\delta_{th}=0.414$.

In order to study statistical properties of peaks in the density field, where PBHs could form, peak theory is used. The number density of the peaks is expressed as $n(r)=\sum_p\delta_D(\mathbf{r}-\mathbf{r_p})$, where $\delta_D$ represents the Dirac delta function and $\mathbf{r_p}$ signifies the position where $\delta$ reaches a local maximum. This requirement of identifying local maxima requires not only the density contrast but also its spatial derivatives upto the second order, leads us into the realm of a ten dimensional joint probability distribution function ${{P}}(\{y_i\})$. This set of variables $\{y_i\}$, density contrasts and its derivatives $y_1=\delta$, $y_2=\partial_1\delta$, $...$, $y_5=\partial_1\partial_1\delta$, $...$, and $y_{10}=\partial_2\partial_3\delta$ allows us to capture the local behavior of the density field around each peak. The density contrast field is assumed as a Gaussian random field and the joint distribution of these variables is also gaussian \cite{Wu2020}.

\begin{equation}
    P(\{y_i\})=\frac{\exp\left({\frac{1}{2}\sum_{ij}\Delta y_i\mathcal{M}_{ij}^{-1}\Delta y_j}\right)}{\sqrt{(2\pi)^{10}}det\mathcal{M}}
\end{equation}
where $\mathcal{M}$ represents the covariance matrix, $\Delta y_i=y-<y_i>$. A sequence of dimensional reductions ultimately lead to the simplification of $P(\{y_i\})$ to $P(\nu)$, a one dimensional conditional probability distribution function \cite{1986ApJ...304...15B}. Using the $P(\nu)$ distribution function, the number density of peaks $n(\nu_{th})$ for cases $\nu>\nu_{th}$ can be expressed in an integral form,

\begin{equation}
    n(\nu_{th})=\frac{1}{(2\pi)^2}\left(\frac{\sigma_2}{\sqrt{3}\sigma_1}\right)^3\int_{\nu_{th}}^\infty G(\gamma,\nu) \exp\left(-\frac{\nu^2}{2}\right)d\nu\; .
\end{equation}
Here the parameter $\gamma=\frac{\sigma_1^2}{\sigma_{\delta}\sigma_2}$ within the function $G(\gamma,\nu)$ encapsulates the characteristics of the $\delta$ profile, bearing essential information about its shape and profile. Hence, the fraction of the PBH mass can be calculated as
\begin{equation}
    \beta(M_{PBH})=\frac{1}{\sqrt{2\pi}}\left(\frac{\mathfrak{R}\sigma_2}{\sqrt{3}\sigma_1}\right)^3\int_{\nu_{th}}^\infty G(\gamma,\nu)\exp\left({-\frac{\nu^2}{2}}\right)d\nu\; .
\end{equation}
Due to the inherent complexity of the $G(\gamma,\nu)$ function, several approximations have been proposed. In particular, Green, Liddle, Malik, and Sasaki (GLMS) proposed the approximation where $\nu \gg 1$ and $\gamma\approx1$, simplifying the situation to only two independent spectral moments: $\sigma_{\delta}$ and $\sigma_1$. Within this GLMS approximation, the expression for $\beta(M_{PBH})$ can be analytically obtained as
\begin{equation}
    \beta(M_{PBH})=\frac{1}{\sqrt{(2\pi)}}\left(\frac{\mathfrak{R}\sigma_1}{\sqrt{3}\sigma_\delta}\right)^3\left(\nu_{th}^2-1\right)\exp\left({-\frac{\nu_{th}^2}{2}}\right)\; .
\end{equation}
\begin{figure}[h!]
\centering
\includegraphics[width=1\textwidth]{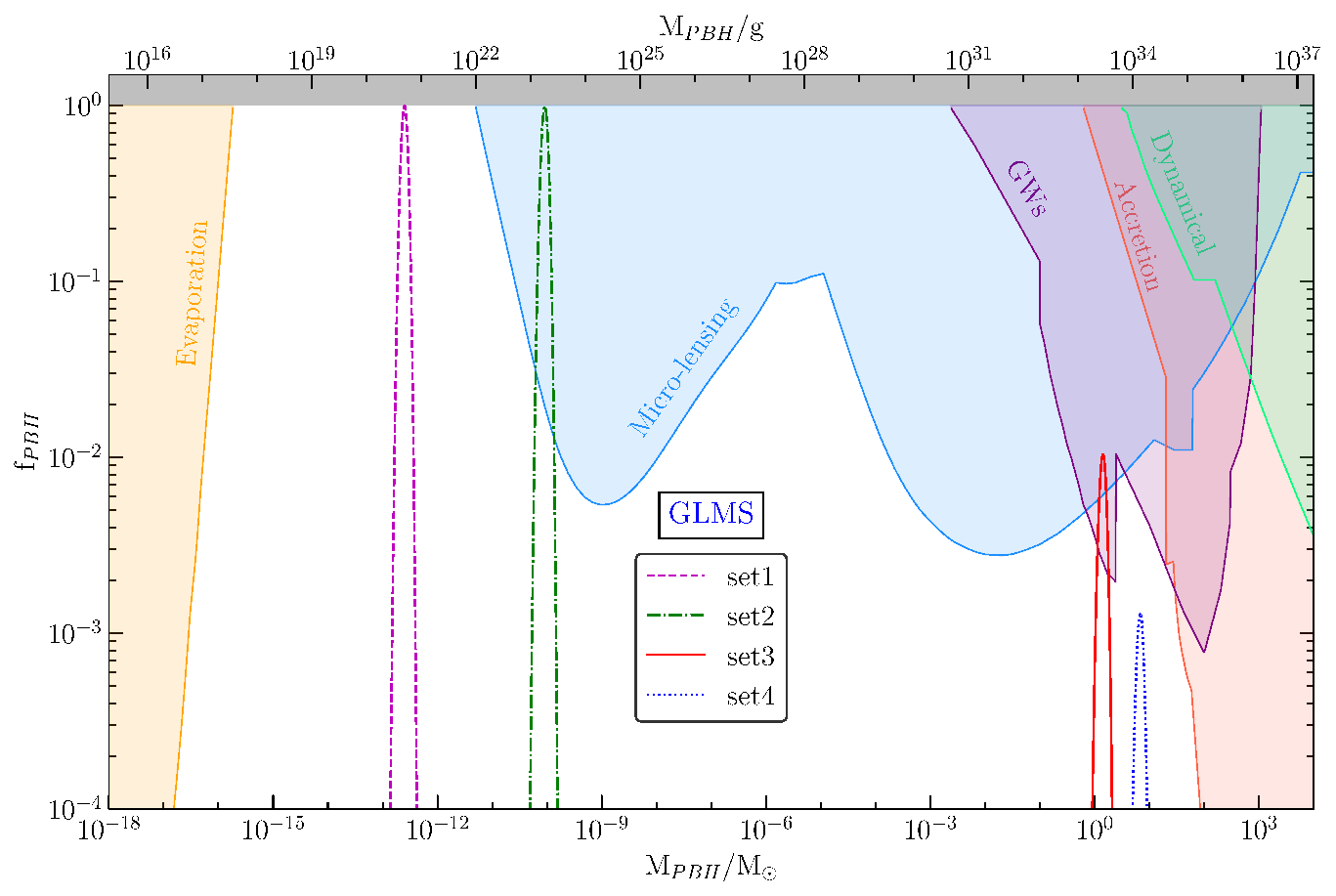}
\caption{Fractional abundance of PBHs in the framework of GLMS approximation in peak theory formalism is plotted as a function of PBH mass. This graphical representation pertains to the four distinctive cases discussed in Table \ref{tab:PBH_mass}. The shaded region represents the constraints on $f_{PBH}$ from various observational limits, including PBH evaporation, microlensing, GWs, PBH accretion and dynamical constraints}
\label{Fig3.4}
\end{figure}

In accordance with the parameter values in Table \ref{tab:PBH_mass}, $f_{PBH}(M_{PBH})$ within our model has been plotted in Figure \ref{Fig3.4} employing the GLMS approximation in peak theory formalism and it is found to be consistent with the recent observational constraints.

\subsection{Press-Schechter formalism}

Now, let us turn our attention to the calculation of $\beta(M_{PBH})$ using PS formalism.  Within the framework of the PS formalism, $\beta(M_{PBH})$ for a specific mass value is defined as the likelihood that the density contrast $\delta$ smoothed over the comoving Hubble scale $\mathfrak{R}=\frac{1}{k_{PBH}}=\frac{1}{(aH)_{PBH}}$ through an appropriate window function exceeds the threshold $\delta_{th}$ required for PBH formation \cite{PhysRevD.88.084051, PhysRevD.100.123524, PhysRevD.101.044022}. 
\begin{equation}
    \beta(M_{PBH})=\gamma\int_{\delta_{th}}^{1}\mathcal{P}(\delta)d\delta
\end{equation}
\begin{equation}\label{beta}
    \beta(M_{PBH})=\gamma\int_{\delta_{th}}^1\frac{d\delta}{\sqrt{2\pi}\sigma_{\delta}}exp\left[-{\frac{\delta^2}{2\sigma^2_{\delta}}}\right]\approx\gamma\frac{\sigma_{\delta}}{\sqrt{2\pi}\delta_{th}}exp\left[-\frac{\delta^2_{th}}{2\sigma^2_{\delta}}\right] .
\end{equation}
The equation (\ref{beta}) implies the dependence of $\beta(M_{PBH})$ on the chosen value of $\delta_{th}$ and for nearly monochromatic PBH mass functions $\delta_{th}=0.414$.

\begin{figure}[h!]
\centering
\includegraphics[width=0.9\textwidth]{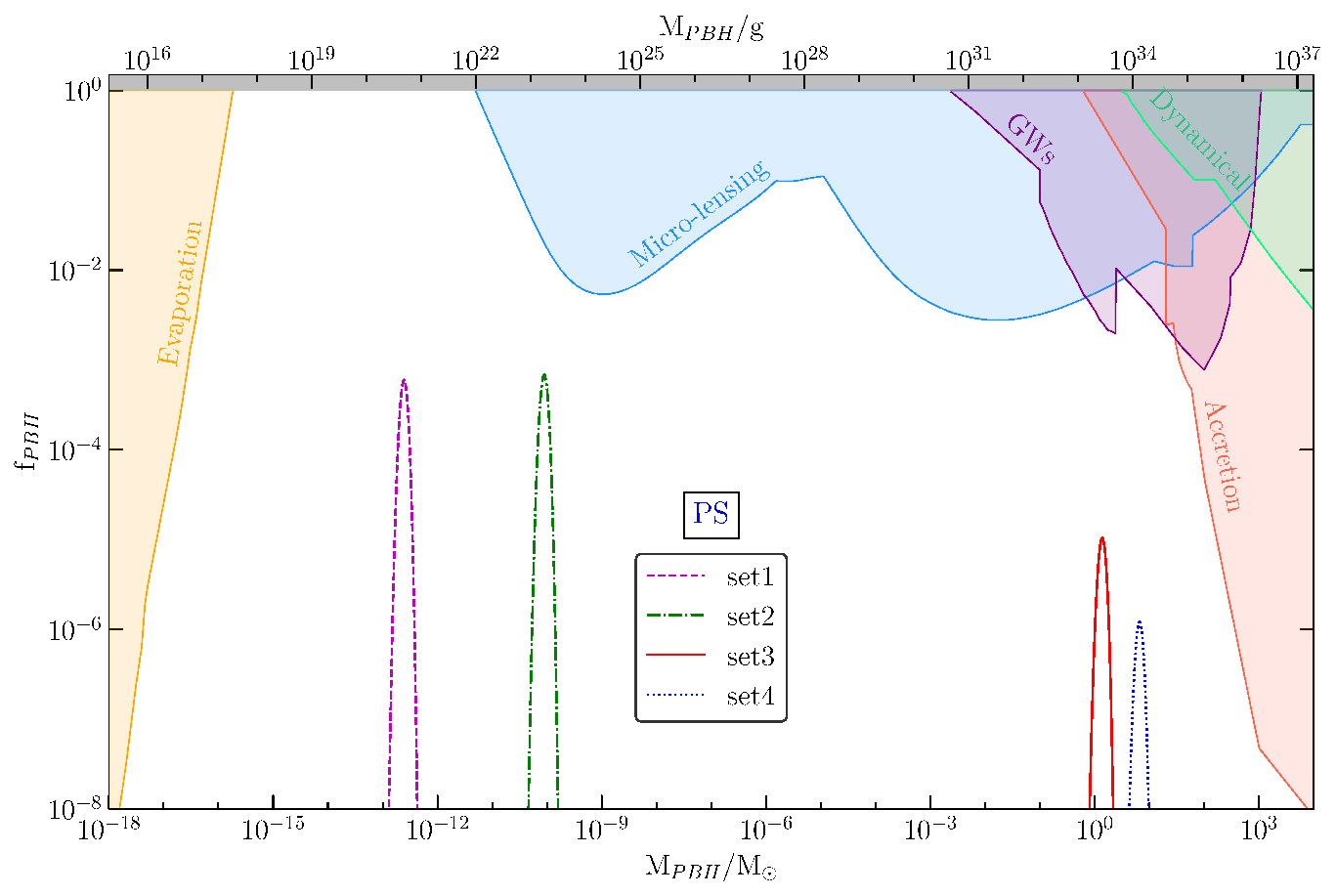}
\caption{Fractional abundance of PBHs in the framework of Press-Schechter formalism is plotted as a function of PBH mass. This graphical representation pertains to the four distinctive cases discussed in Table \ref{tab:PBH_mass}.  The shaded region represents the constraints on $f_{PBH}$ from various observational limits, including PBH evaporation, microlensing, GWs, PBH accretion and dynamical constraints} 
\label{Fig3.5}
\end{figure}

Substituting equation (\ref{sigma2}) in (\ref{beta}), one can compute $\beta(M_{PBH})$ within the framework of PS formalism and hence the $f_{PBH}(M_{PBH})$. In accordance with the parameter values in Table \ref{tab:PBH_mass}, $f_{PBH}(M_{PBH})$ within our model has been plotted in Figure \ref{Fig3.5} employing the PS formalism.

\begin{figure}[h!]
\centering
\includegraphics[width=0.9\textwidth]{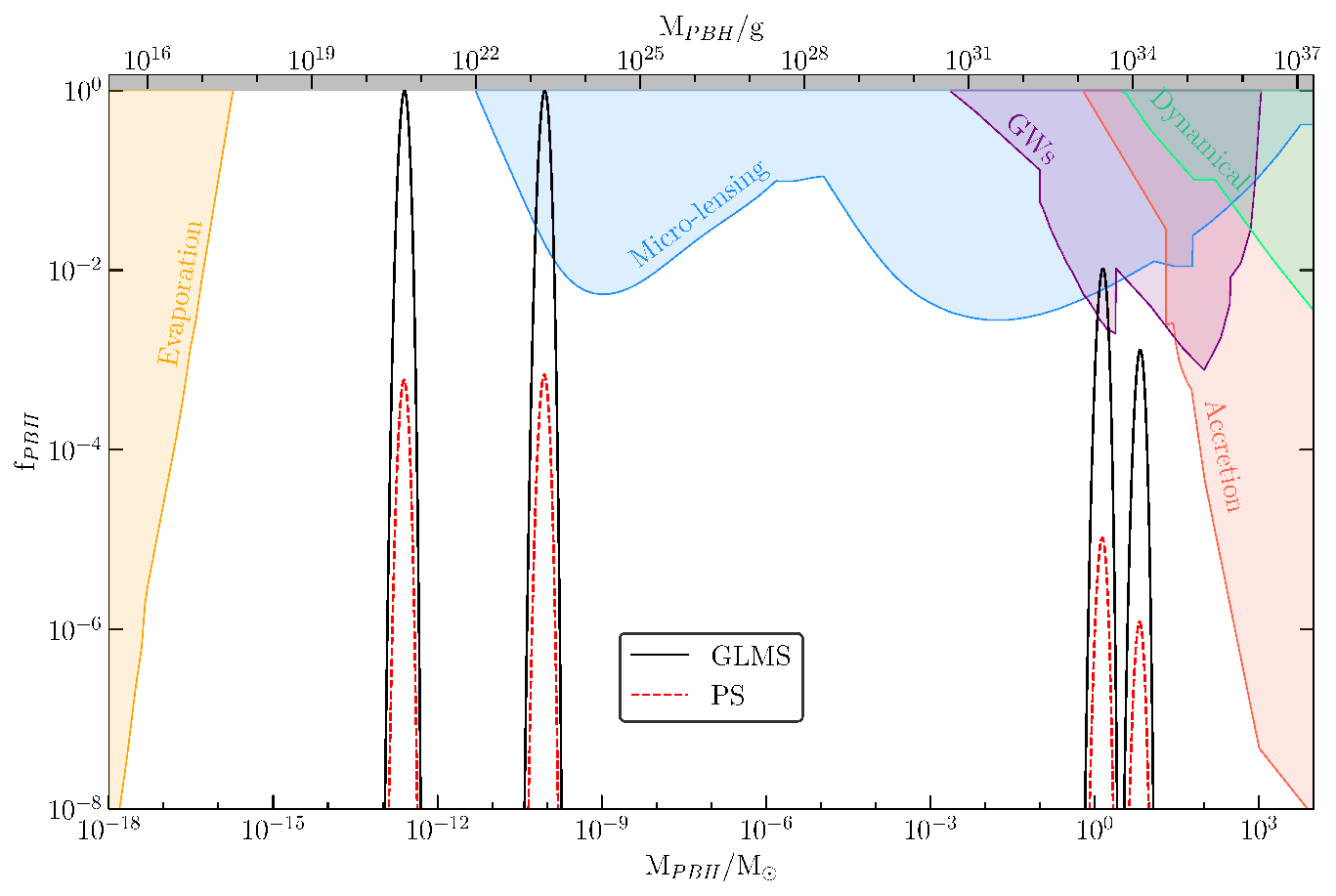}
\caption{Comparative analysis of the fractional abundance of PBHs across four different PBH mass scales as in Table \ref{tab:PBH_mass}  in the framework of GLMS perspective and PS formalism. The shaded region represents the constraints on $f_{PBH}$ from various observational limits, including PBH evaporation, microlensing, GWs, PBH accretion, and dynamical constraints} 
\label{Fig3.6}
\end{figure}

In addition we compare the PBH masses in four typical mass windows with the PBH abundances derived from the GLMS approximation and the PS theory. It is observed that the PS theory consistently underestimates $f_{PBH}$ by a significant margin, typically by two to three orders of magnitude in comparison to peak theory \cite{PhysRevD.104.083546}, and it is shown in Figure \ref{Fig3.6}.

\section{Conclusions}
PBHs have the potential to have a highly significant impact on various astrophysical and cosmological phenomena. PBHs make a substantial contribution to the current density of dark matter in the universe. Furthermore, PBHs may serve as seeds for the emergence of supermassive blackholes and give rise to the formation of binary blackhole systems, which are pertinent to the detection of gravitational waves by observatories like LIGO and Virgo. In the framework of the inflationary paradigm, formation of PBHs offers a captivating realm for research. In these models large modes of fluctuation that exit the Hubble radius during inflation can result in the PBH formation when they renter it during the radiation dominated era. In a typical single-field inflation model, the presence of a feature resembling a near-inflection point can significantly enhance the primordial fluctuations to several orders of magnitude leading to PBH formation.

In this work we consider an inflationary model that incorporates a sharp step on a chaotic potential. The sharp step functions as a deceleration mechanism for the inflaton, causing it to reduce its speed. This inturn significantly increases the amplitude of scalar perturbations. It is intriguing to observe that, for the cosmological abundances of PBHs formation to occur, the primordial scalar power spectrum needs to be enhanced to the order of $10^{-2}$. The straightforward nature of our potential enables the separate consideration of observables such as $n_s$ and $r$ on the CMB scale and at small scales where one expects PBHs formation. As a result, our model can account for the formation of PBHs on a small scale in four distinct mass ranges, ranging from an incredibly small $10^{-13}M_\odot$ to $6M_{\odot}$. When adjusting the parameters, as $\phi_{step}$ increases, the peak of the scalar power spectrum $\mathcal{P}(k)$ shifts to larger scales resulting in an increase in the mass of PBHs without affecting the values of $n_s$ and $r$ on the CMB scale. It is worth noting that the conventional approach for determining the fractional abundances of PBHs at present, as employed in our study, relies on the assumption that the mass of PBHs remains constant up to the present era. We systematically compute the fractional abundance of PBHs using GLMS approximation within the framework of peak theory and using PS formalism. Using the GLMS approximation of peak theory, the formation of PBHs with masses $10^{-13}M_\odot$ and $10^{-10}M_\odot$ results in $f_{PBH}\approx1$ respectively. This can contribute to all dark matter alone. The PBHs with masses $1M_\odot$ and $10M_\odot$ result in $f_{PBH}\approx0.01$ and $0.001$ respectively. Observations of black hole mergers by the LIGO and Virgo collaborations constrain the abundances of PBHs to be $f_{PBH} \leq 0.01$ within the mass range of 1 to 300$M_\odot$. The fractional abundance of PBHs determined is in alignment with this observational constraints.

Also it is noted  that the PS theory significantly underestimates the $f_{PBH}$ of PBHs by a factor of two to three orders of magnitude. This underestimation occurs because the PS theory simplifies the more comprehensive peak theory, leading to a systematic bias, especially in the context of USR inflation. In general, the peak theory is more solidly grounded, and the PS theory should be seen as its simplified version, so we need to be careful when interpreting the results. In conclusion this article is focused on exploring the formation and $f_{PBH}$ of PBHs through the use of the GLMS perspective of peak theory, with a specific focus on chaotic potential with a sharp step. Also this step feature in the inflationary potential step allows the decoupling of CMB-scale observables from small-scale physics, facilitating the PBH formation as a dark matter candidate.

\section*{Acknowledgments}

MJ acknowledges the Associateship of IUCAA. We thank the reviewers for the valuable comments and suggestions, which helped to improve the quality of our work.

\end{document}